\documentclass[runningheads]{llncs}
\usepackage[T1]{fontenc}
\usepackage{graphicx}
\usepackage{multirow}
\usepackage{subcaption}
\begin{document}
\title{AutoPET Challenge: Tumour Synthesis for Data Augmentation}
\author{Lap Yan Lennon Chan\inst{1} \and
Chenxin Li\inst{2} \and
Yixuan Yuan\inst{2}}
\authorrunning{L. Chan et al.}
\institute{Department of Computer Science and Engineering, The Chinese University of Hong Kong, Shatin, Hong Kong SAR \and
Department of Electronic Engineering, The Chinese University of Hong Kong, Shatin, Hong Kong SAR}
\maketitle              % typeset the header of the contribution
\begin{abstract}
% In this paper, we explore the feasibility of utilising a deep generative model to augment a limited dataset for the task of automated lesion segmentation in whole-body PET/CT scans. The small size of the dataset is a known obstacle to achieving high performance in lesion segmentation models. Our approach leverages a deep generative model to synthesize new PET-CT images with lesions, expanding the dataset and potentially improving the generalizability of the segmentation model. We adapt the DiffTumor method \cite{difftumour}, which utilizes healthy CT images and lesion segmentation masks to generate synthetic CT images containing tumours for our use. Our adaptation modifies DiffTumor to accommodate PET-CT images. We train the deep generative model on the AutoPET dataset and employ it to generate synthetic PET-CT images with lesions. We then evaluate the performance of a segmentation model trained on the augmented dataset in comparison to a model trained on the original dataset. Our findings demonstrate that the segmentation model trained on the augmented dataset achieves a higher Dice score than the model trained on the original dataset, suggesting the effectiveness of our data augmentation approach in this application. Further analysis is necessary to fully understand the strengths and limitations of our method, but this work presents a promising direction for improving lesion segmentation in whole-body PET/CT scans with limited datasets.

Accurate lesion segmentation in whole-body PET/CT scans is crucial for cancer diagnosis and treatment planning, but limited datasets often hinder the performance of automated segmentation models. In this paper, we explore the potential of leveraging the deep prior from a generative model to serve as a data augmenter for automated lesion segmentation in PET/CT scans. We adapt the DiffTumor method, originally designed for CT images, to generate synthetic PET-CT images with lesions. Our approach trains the generative model on the AutoPET dataset and uses it to expand the training data. We then compare the performance of segmentation models trained on the original and augmented datasets. Our findings show that the model trained on the augmented dataset achieves a higher Dice score, demonstrating the potential of our data augmentation approach. In a nutshell, this work presents a promising direction for improving lesion segmentation in whole-body PET/CT scans with limited datasets, potentially enhancing the accuracy and reliability of cancer diagnostics.

\keywords{Tumour Generation  \and Data Augmentation \and PET-CT}
\end{abstract}
\section{Introduction}
Over the past decades, PET/CT has emerged as a pivotal tool in oncological diagnostics, management and treatment planning~\cite{heron2008pet}. Lesion segmentation, which is a necessary step for quantitative image analysis, is performed manually and is thus tedious, time-consuming and costly~\cite{sun2022few,ding2022unsupervised}. Machine Learning, however, offers the potential for fast and fully automated quantitative analysis of PET/CT images. With this in mind, we participated in the AutoPET-III challenge held in MICCAI 2024, the task of which was to automate the Lesion Segmentation of Whole-Body PET/CT \cite{Autopet_III}. We adopted a data-centric approach because of the inherent imperfections in the given dataset by the organiser. In this approach, we intended to perform data augmentation to the AutoPET dataset \cite{Autopet_I_dataset,Autopet_III_dataset} without changing the baseline segmentation model (which in this challenge is DynUnet, a MONAI implementation of nnUnet) \cite{DynUNet,nnU_Net}. Our intended approach can best be summarised as utilizing deep generative models~\cite{croitoru2023diffusion,li2024endora,li2024u} to augment the imperfect AutoPET dataset such that the baseline segmentation model achieves better generalization ability. 
\subsection{Exploratory Data Analysis }
\subsubsection{Insufficient dataset size}
\begin{figure}
    \centering
    \includegraphics[width=0.5\linewidth]{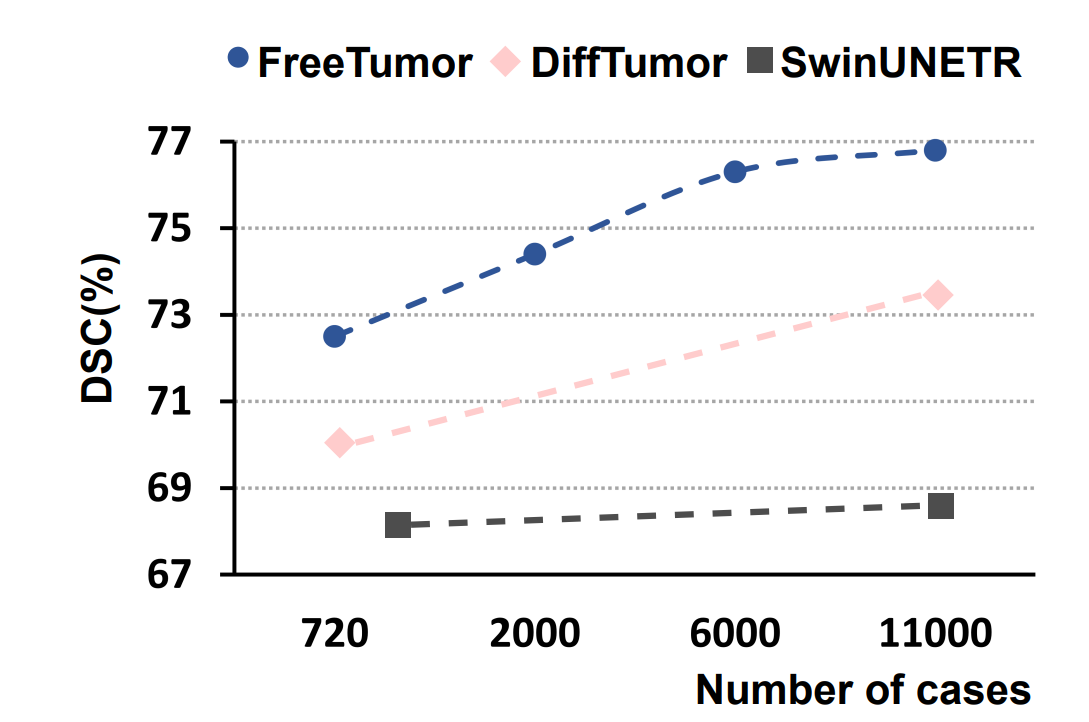}
    \caption{Dice score for tumour segmentation models trained with different numbers of cases (real or synthesized), the trend shows that in general, more training cases lead to better performance. The result of DiffTumor \cite{difftumour} on 11k is re-implemented.}
    \label{fig:scaling_law_freetumor}
\end{figure}

The dataset contains 1614 paired CT-PET scans and lesion segmentation masks. However, previous research \cite{freetumour} shows how the size of the training dataset corresponds positively with the performance of the tumour segmentation model (Figure~\ref{fig:scaling_law_freetumor}). And the training dataset’s size of 1614 could be a limiting factor in segmentation model performance.
\subsubsection{Countermeasures} To address the problem of a small dataset, our intuition is to generate synthesized PET-CT images to augment the dataset, such that we could enlarge the dataset while introducing more diverse, but still in-distribution data into the training set for more robust learning.

\section{Related Work}
\subsection{Towards Generalizable Tumour Synthesis (DiffTumor) \cite{difftumour}}
The researchers propose the synthesis of tumoured CT images from healthy CT images using a lesion segmentation mask as a condition for augmenting a dataset of tumored CT images to train a more robust tumour segmentation model.

The whole pipeline consists of 3 steps. In step 1, healthy CT images are used to train an autoencoder that represents CT images in a compressed latent space. In step 2, a latent diffusion model is trained to generate the latent representation of tumoured CT images with the condition of healthy CT images and the tumour segmentation mask (along with the organ segmentation mask) by learning the reverse process of adding noises to the latent representation of the tumoured CT images. In Step 3, tumoured CT images are synthesized with out-of-dataset healthy CT images and a tumour segmentation mask (again, along with the organ segmentation mask) to augment the original dataset, which is then used to train a tumour segmentation model that performs much better than if it is trained by a dataset that is not augmented.

A better performance with synthetic tumours than with real tumours is achieved under a nnU-Net \cite{nnU_Net} backbone (Table~\ref{table:difftumour performance}).

\begin{table}[h!]
\centering
\begin{tabular}{lccccccc}
\textbf{nnU-Net} & metrics & fold0 & fold1 & fold2 & fold3 & fold4 & average \\
\hline
\multirow{2}{*}{real tumours} & DSC (\%) & 73.8 & 76.8 & 80.0 & 80.5 & 73.4 & 76.9 \\
& NSD (\%) & 62.7 & 70.2 & 71.2 & 70.8 & 67.5 & 68.5\\
\hline
\multirow{2}{*}{DiffTumor} & DSC (\%) & 84.5 & 83.4 & 81.6 &83.9 & 77.3 & 82.1 \\
& NSD (\%) & 78.3 & 74.4 & 74.1 & 76.9 & 72.3 & 75.2\\
\hline
\end{tabular}
\caption{Performance of nnU-Net on different metrics}
\label{table:difftumour performance}
\end{table}

\section{Methodology}
We adapted DiffTumor to our task to conduct data augmentation, then compared the performance of DynUnet trained on the augmented dataset and that of the baseline DynUnet.
\subsection{DynUnet baseline}
We use the DynUnet \cite{DynUNet,nnU_Net}, which is based on nnU-Net. The nnU-Net fingerprint extractor and planner on the autoPET3 dataset are utilized to configure the model.
\subsection{DiffTumor adaptation}
The pipeline for data augmentation using a deep generative model borrows largely from DiffTumor {difftumour} but we adapt it to our use. Most notably, instead of taking in and outputting CT images, PET-CT images are taken in and outputted. In Step 1, whereas DiffTumor trains its autoencoder on a large healthy CT dataset, we train ours only with the AutoPET dataset following the principle of data augmentation, where no new external data should be introduced. In Step 2, unlike DiffTumor which trains only on a certain organ as it wants to demonstrate the generalizability to other organs, we do not need to or intend to demonstrate that. Therefore, Step 2 is trained on the whole Autopet dataset. For Step 2, since we do not have the ground truth organ segmentation mask to accompany the tumour segmentation mask, we utilized SegResNet from “Wholebody ct segmentation” model provided in MONAI model zoo \cite{totalsegmentator,HECKTOR,SegResNet} to produce a pseudo-organ mask instead. We do not follow Step 3 of DiffTumor since we will use the training settings of DynUnet \cite{DynUNet,ROI}.

\section{Results}
\subsection{Experimental Setting}
The AutoPET dataset contains 1614 full-body PET-CT scans, 1291 of which belong to the training set and 323 belong to the validation set. In our experiment, we use the 1291 training data to train the deep generative model for data augmentation. During data augmentation, we generate 3 different synthesized data per 1 training data, thus the augmented training set has a size of 5152 (with some synthesized data from 4 images rejected due to failure of generating pseudo-organ mask by SegResNet).

The input patch size is configured to be (128, 160, 112) with a batch size of 2. Training runs for a total of 581 epochs. 

\subsection{Quantitative Result}

Table~\ref{table:our performance} shows the dice scores for our trained segmentation model (DynUnet) on the baseline dataset and the augmented dataset under different data transformation techniques (and thus different dataset size). We have completed experiments both under the condition of taking 1 random transform per image and 15 random transforms per image.

\begin{table}[h!]
\centering
\begin{tabular}{cccc}
Transforms per image & Baseline/Augmented & Training Dataset size & DSC \\
\hline
\multirow{2}{*}{1} & Baseline & 1291 & 0.3650 \\
& Augmented & 5152 & 0.4542\\
\hline
\multirow{2}{*}{15} & Baseline & 19365 & 0.5398 \\
& Augmented & 77280 & 0.6143\\
\hline
\multirow{2}{*}{30} & Baseline & 38730 & 0.5859 \\
& Augmented & 154560 & 0.6179\\
\hline

\end{tabular}
\caption{Quantitative Comparisons between the Baseline Model and Model trained on the Augmented Dataset}
\label{table:our performance}
\end{table}

\subsection{Qualitative Results}
\subsubsection{Predicted Mask}
\begin{figure}
    \centering
    \includegraphics[width=1\linewidth]{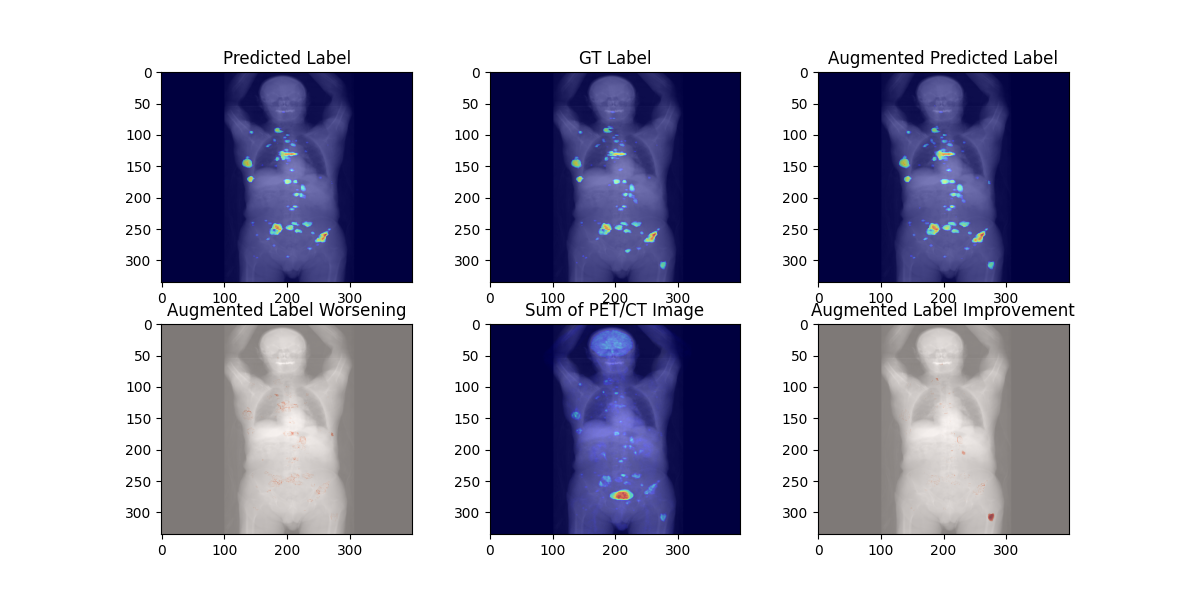}
    \includegraphics[width=1\linewidth]{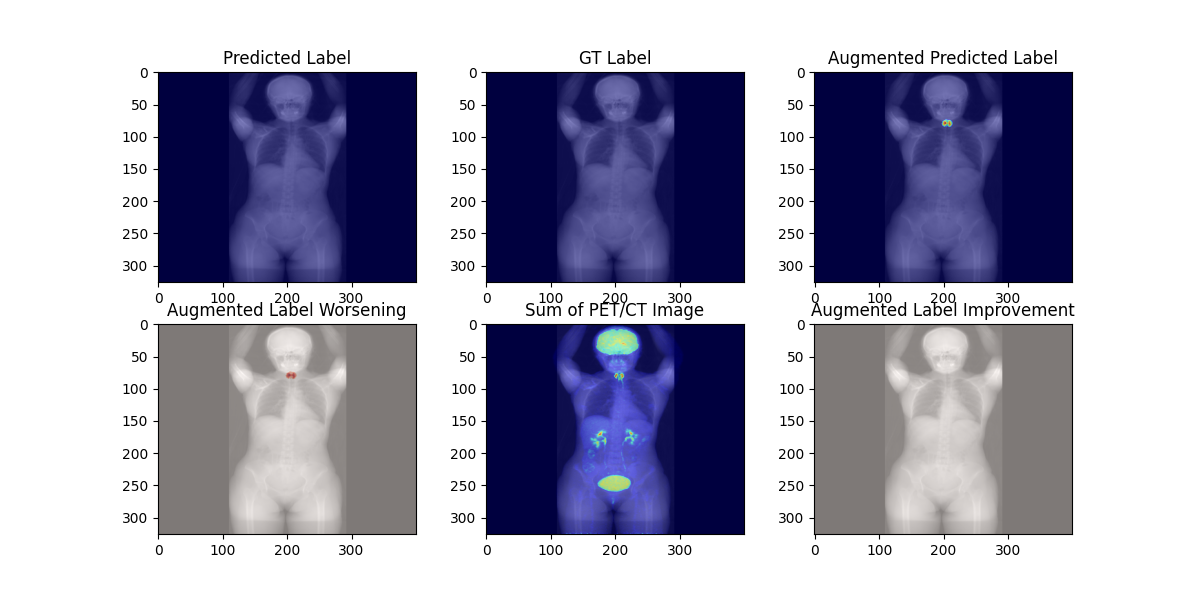}
    \caption{The successful and unsuccessful case of lesion segmentation}
    \label{fig:pos-neg}
\end{figure}

The following (Figure~\ref{fig:pos-neg}) are two examples taken from the validation set. The first one is a success in that it improves from the baseline model by successfully segmenting a tumour which the baseline fails to do. The second one, however, presents a false positive displayed by the model trained on the augmented dataset not displayed by the baseline. All in all, while in general, the model trained on the augmented dataset improves from the baseline, there are cases where its performance worsens just as there are cases where it improves. 

\begin{figure}
    \centering
    \includegraphics[width=\linewidth]{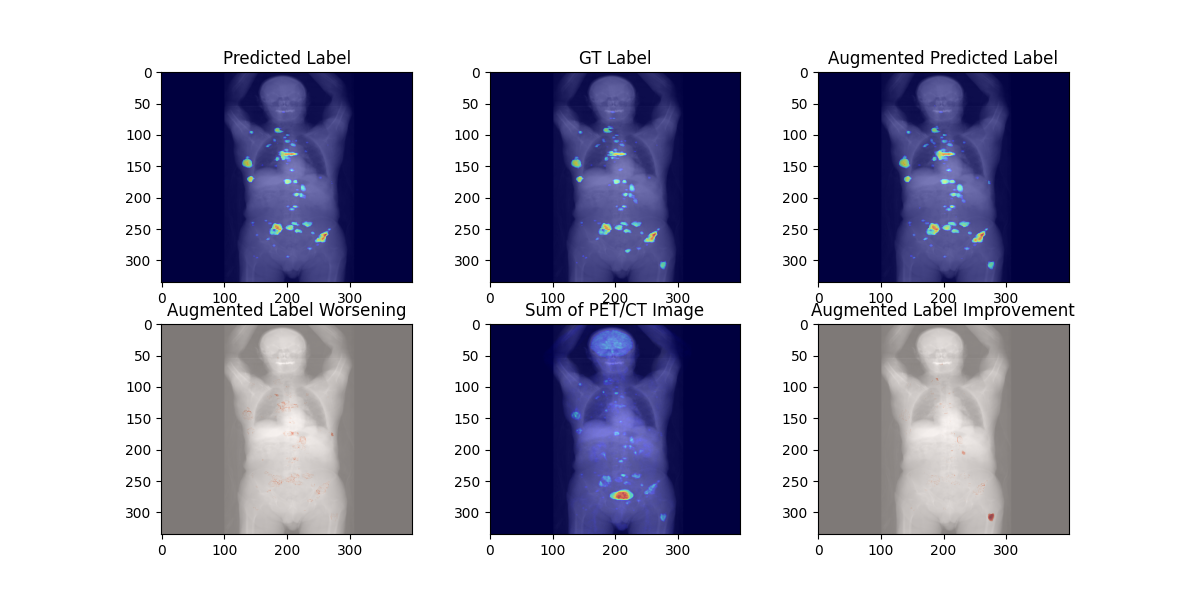}
    \includegraphics[width=\linewidth]{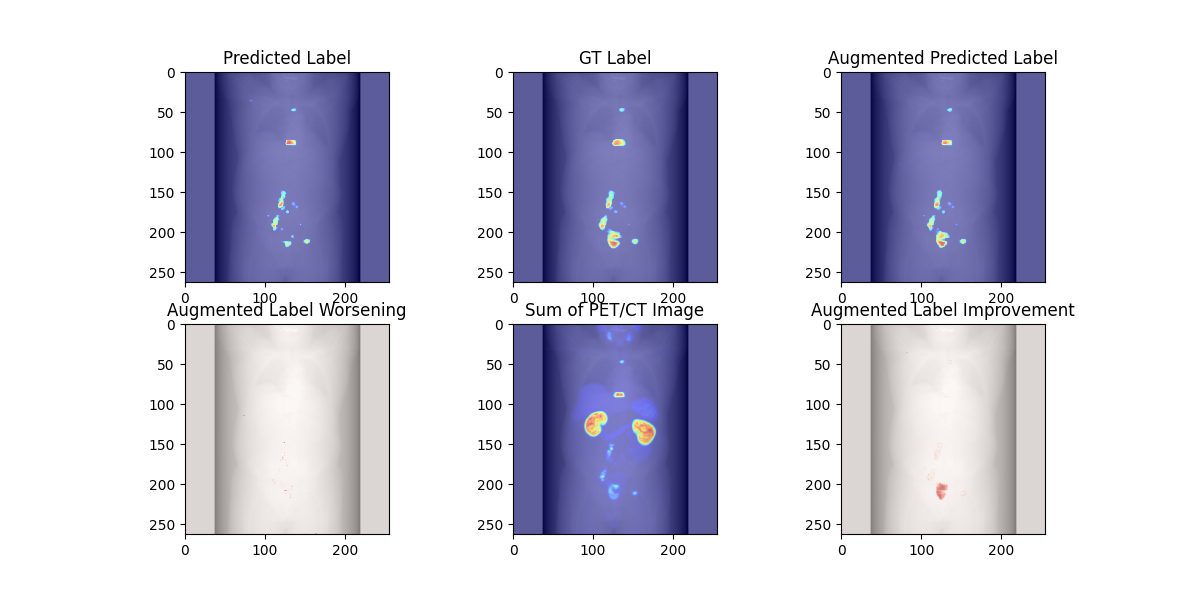}
    \caption{More Lesion Segmentation Results}
    \label{fig:more_examples_2}
\end{figure}

\noindent Below are some more examples (Figure~\ref{fig:more_examples} and ~\ref{fig:more_examples_2}).:

\begin{figure}
    \centering
    \includegraphics[width=\linewidth]{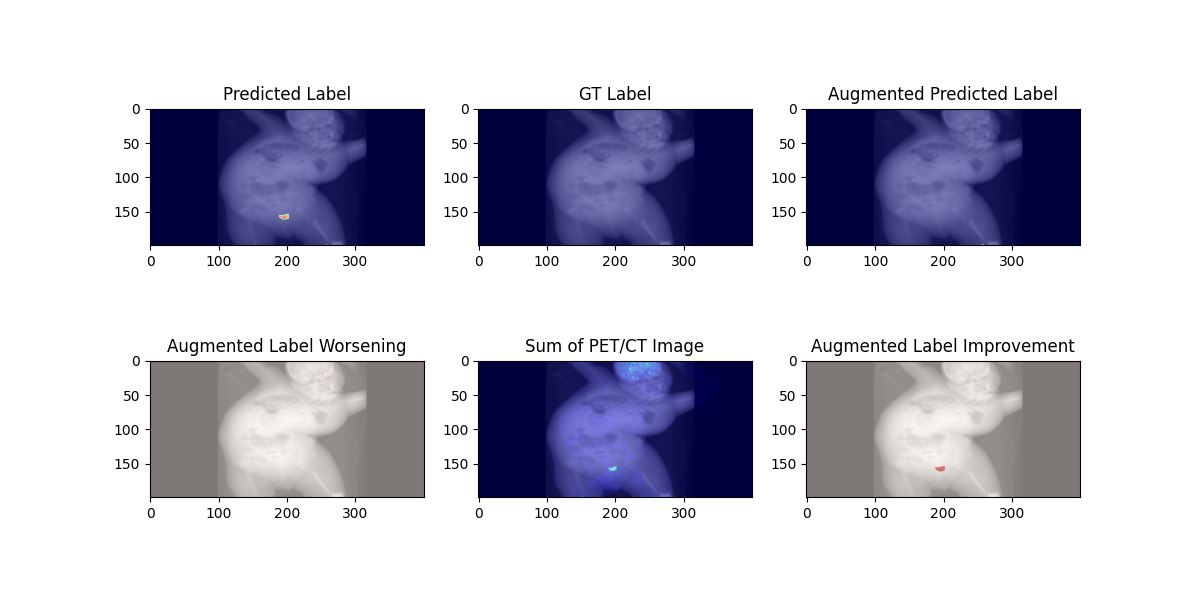}
    \includegraphics[width=\linewidth]{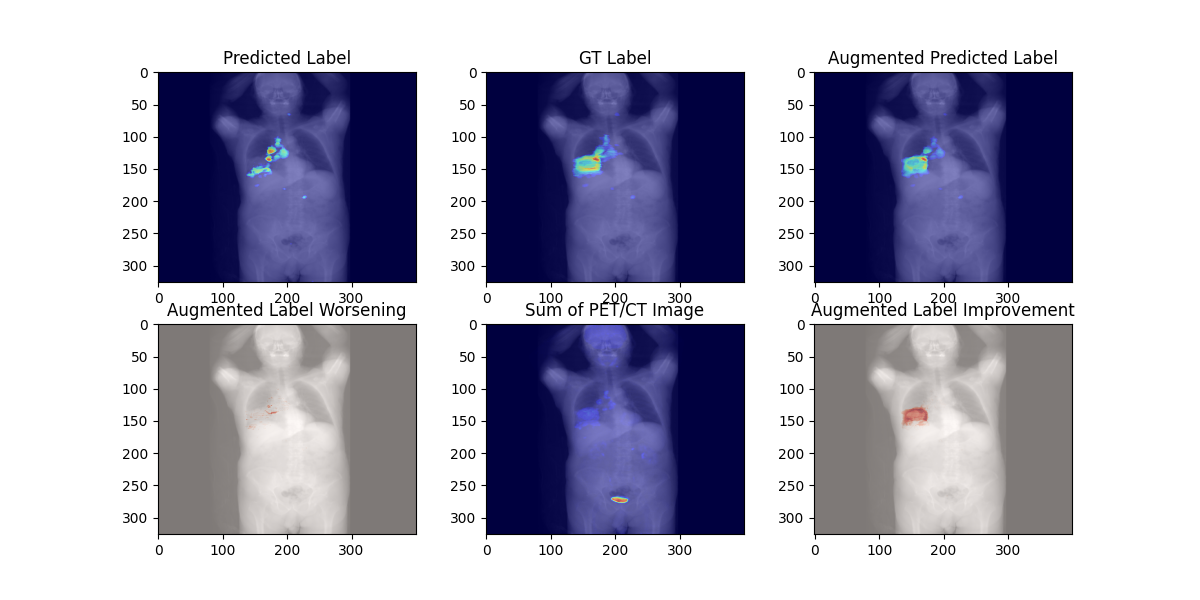}
    \caption{More Lesion Segmentation Results}
    \label{fig:more_examples}
\end{figure}

\subsubsection{Synthesized Tumour}
Here are some tumours synthesized on the validation set (Figure~\ref{fig:synthesized}):

\begin{figure}
    \centering
    \includegraphics[width=1\linewidth]{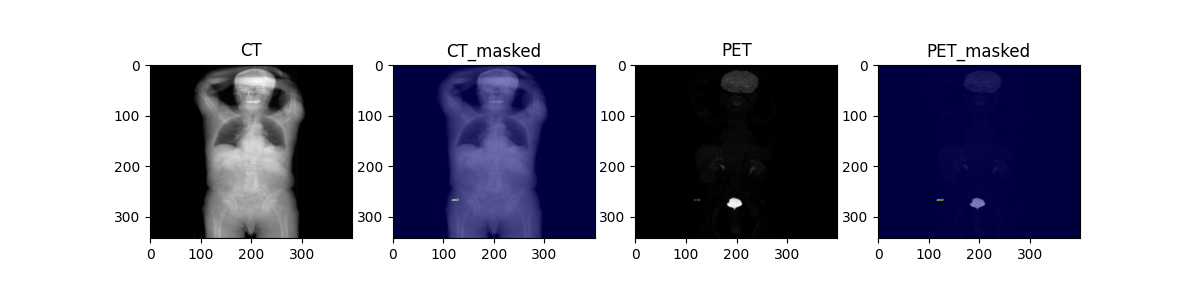}
    \includegraphics[width=1\linewidth]{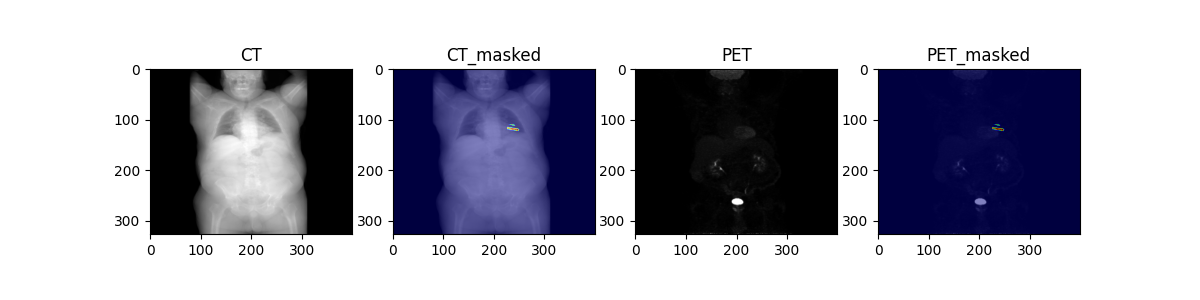}
    \caption{Tumour Generation in unseen data. The generation network learns well by not automatically translating every tumour to be shown as a PET tracer as that is not always the case.}
    \label{fig:synthesized}
\end{figure}

\section{Conclusion}
% In conclusion, we demonstrated the feasibility of this new paradigm of data augmentation through using Deep Generative Model to learn the distribution of the existing data and then synthesize data to augment the original dataset for downstream application in the task of multitracer multicentre full-body PET-CT lesion segmentation. However, further analysis is to be carried out to fully understand its benefits and downfalls, as well as the extent it can push forward the state-of-the-art in this task. Below are some suggestions to further improve performance as well as to more thoroughly evaluate our method.

In conclusion, we have successfully demonstrated the viability of a novel data augmentation paradigm for multitracer, multicentre full-body PET-CT lesion segmentation. This approach leverages a Deep Generative Model to learn the distribution of existing data and synthesize new samples, thereby augmenting the original dataset for downstream applications. Our results show promise in advancing the state-of-the-art in this challenging task. However, we acknowledge that further comprehensive analysis is necessary to fully elucidate the benefits and potential limitations of this method.

\begin{credits}
\subsubsection{\ackname} This study was funded by the Faculty of Engineering of the Chinese University of Hong Kong. The authors acknowledge the AutoPET challenge for the free publicly available PET/CT images used in this study.

\subsubsection{\discintname}
The authors have no competing interests to declare that are relevant to the content of this article. 
\end{credits}

\end{document}